\documentclass[%
reprint,
superscriptaddress,
showpacs,preprintnumbers,
amsmath,amssymb,
aps,
showkeys,
prb,
notitlepage
]{revtex4-1}

\usepackage{graphicx}
\usepackage[export]{adjustbox}
\usepackage{dcolumn}
\usepackage{bm}
\usepackage{wasysym}
\usepackage[utf8]{inputenc}
\usepackage{multirow}
\usepackage{xcolor}

\begin{document}

\title{Boron and nitrogen doping in graphene antidot lattices}


\author{S\o ren J. Brun}%
\affiliation{Department of Physics and Nanotechnology, Aalborg University, DK-9220 Aalborg \O st, Denmark}
\affiliation{Center for Nanostructured Graphene (CNG), DK-9220 Aalborg \O st, Denmark}
\author{Vitor M. Pereira}%
\affiliation{Centre for Advanced 2D Materials and Department of Physics, National University of Singapore, 2 Science Drive 3, Singapore 117542}
\author{Thomas G. Pedersen}%
\affiliation{Department of Physics and Nanotechnology, Aalborg University, DK-9220 Aalborg \O st, Denmark}
\affiliation{Center for Nanostructured Graphene (CNG), DK-9220 Aalborg \O st, Denmark}

\begin{abstract}

Bottom-up fabrication of graphene antidot lattices (GALs) has previously yielded atomically precise structures with sub-nanometer periodicity. Focusing on this type of experimentally realized GAL, we perform density functional theory calculations on the pristine structure as well as GALs with edge carbon atoms substituted with boron or nitrogen. We show that p- and n-type doping levels emerge with activation energies that depend on the level of hydrogenation at the impurity. Furthermore, a tight-binding parameterization together with a Green's function method are used to describe more dilute doping. 

\end{abstract}

\maketitle


\section{Introduction}

Since its discovery,\cite{novoselov2004electric} graphene has shown many interesting properties such as ultra-high electron mobility,\cite{novoselov2004electric,novoselov2005two,bolotin2008ultrahigh} high transparency\cite{nair2008fine} and record-breaking mechanical strength.\cite{lee2008measurement} However, one major drawback is the lack of a band gap, which is required for obtaining high on/off ratios in field-effect transistors.\cite{huang2007making} Therefore, immense effort has been put into turning graphene into a semiconductor while preserving as much as possible its intrinsic characteristics. A popular method is dimensional narrowing, forming graphene nanoribbons (GNRs), which has been shown to introduce a tunable band gap dependent on the width and chirality.\cite{obradovic2006analysis,son2006energy,han2007energy} Another promising and widely studied method is to periodically alter graphene in two dimensions. Hydrogen adsorption onto graphene on an iridium surface has been shown experimentally to create a periodic pattern and open a band gap.\cite{balog2010bandgap} Periodic gating has been studied as well, but was found to not open a band gap large enough for practical applications.\cite{pedersen2012band} Finally, graphene antidot lattices (GAL) can be defined by means of periodic two-dimensional patterning in the form of perforations, which opens a widely tunable band gap depending on the geometry, characteristic dimensions, and chirality that define each element (unit cell) of these superlattices.\cite{pedersen2008graphene} 

The above-mentioned methods for opening a band gap have been studied experimentally to a great extent using top-down methods.\cite{eroms2009weak,bai2010graphene,liang2010formation,kim2010fabrication,giesbers2012charge,oberhuber2013weak} However, fabricating GNRs along this route can lead to scattering from edge imperfections, which has been shown to degrade the transport properties.\cite{areshkin2007ballistic,mucciolo2009conductance} GNRs may also be fabricated by unzipping carbon nanotubes which leads to much more regular edges.\cite{kosynkin2009longitudinal,jiao2009narrow} Electron-beam lithography has been utilized to create GALs with periods down to a few dozen of nanometers, and experimentally determined gaps as high as 102 meV have been reported.\cite{kim2010fabrication} However, GALs suffer from the same problems as GNRs when fabricated using top-down methods. The structures lack full periodicity and imperfections lead to scattering. Calculations have shown that disorder is detrimental to the electronic properties of GALs, as the band gap vanishes or is significantly lowered.\cite{yuan2013electronic} Transport calculations support this finding and show that leakage currents may form through disordered graphene antidot devices.\cite{power2014electronic} 

A promising method that can overcome the problems of disorder is to use bottom-up self-assembly for fabrication, which provides much better control of the formed structures. However, research utilizing bottom-up methods to fabricate graphene nanostructures is still in its infancy. Nonetheless, several groups have successfully synthesized various atomically precise nanostructures using such methods. Cai \textit{et al.}\cite{cai2010atomically} have fabricated GNRs and chevron-shaped GNRs, so-called graphene nanowiggles (GNWs), using surface-assisted coupling of two different precursors on an Au(111) surface followed by cyclodehydrogenation. This yielded narrow, fairly long GNRs and GNWs on the surface. Modified versions of the GNW precursor with pyridine-like nitrogen at one or two sites has been used by Bronner \textit{et al.}\cite{bronner2013aligning} to fabricate doped GNWs. Later, Cai \textit{et al.}\cite{cai2014graphene} used these precursors to fabricate GNW heterojunctions and heterostructures by changing between pristine and doped precursors during synthesis. These structures were recently studied theoretically by Lherbier \textit{et al.},\cite{lherbier2015charge} who reported reasonably high mobilities as well as charge carrier separation. Two-dimensional structures have also been prepared using bottom-up procedures. A nitrogenated GAL with C$_2$N stoichiometry has been synthesized by Mahmood \textit{et al.}\cite{mahmood2015nitrogenated} via a wet-chemical technique, producing a network of aromatic rings with nitrogen between them, where they measured a band gap of 1.96~eV. Sánchez-Sánchez \textit{et al.}\cite{sanchez2015surface} utilized cyclodehydrogenation to produce BN-substituted heteroaromatic networks from another precursor. Finally, Bieri \textit{et al.}\cite{bieri2009porous} have used the precursor hexaiodo-substituted macrocycle cyclohexa-m-phenylene (CHP) to produce a GAL on an Ag(111) surface with sub-nanometer periodicity. These new results on bottom-up techniques for producing atomically precise and even doped graphene structures bring hope that graphene could be used for semiconductor electronics. Despite the high level of control on the atomic scale, these methods have some drawbacks as well. The domain size is still limited, and the synthesis takes place on metallic surfaces, requiring the structures to be transferred after fabrication. However, further optimization of the synthesis could improve the structures for device feasibility. 

As mentioned above, doping has been actively pursued in graphene nanostructures in order to fabricate e.g.\ junctions for device application. Usual dopants are boron and nitrogen, as they fit in the lattice easily, but other types of doping have also been studied, such as aluminum, sulfur and phosphorus.\cite{terrones2012role} Nitrogen doped graphene has been synthesized by several groups from methods such as chemical vapor deposition (CVD) on copper using methane and ammonia\cite{wei2009synthesis} or CVD on nickel using triazine.\cite{usachov2011nitrogen} The transport properties of boron or nitrogen doped graphene were studied theoretically by Lherbier \textit{et al.}\cite{lherbier2008charge} while the effect of unbalanced sublattice nitrogen doping was studied by Lherbier and other coworkers.\cite{lherbier2013electronic} Isolated boron and nitrogen doping in GNRs and near graphene edges has also been studied theoretically.\cite{pedersen2013self,pedersen2015self} Nitrogen doped carbon nanotubes\cite{czerw2001identification} and GNRs\cite{wang2009n} have been realized experimentally, and doped GNRs have been studied theoretically to a large extent.\cite{huang2007making,yan2007intrinsic,li2009spin,wang2011selective} It was shown that the most stable configuration of boron and nitrogen doping is at the edges of the nanoribbon and that nitrogen doping can be either pyridine- or pyridinium-like. Scanning Raman spectroscopy has indicated p-type doping in GALs after fabrication from electron-beam lithography and oxygen reactive ion etching.\cite{heydrich2010scanning} These GALs were similar to other top-down fabricated ones, and it was suggested that the doping stems from the patterning process. 

In this paper, we study the effect of introducing doping in the GAL synthesized by Bieri \textit{et al.}\cite{bieri2009porous} in the form of boron or nitrogen impurities. As pointed out by Sánchez-Sánchez \textit{et al.},\cite{sanchez2015surface} the method of cyclodehydrogenation may be extended to more complex systems, provided the precursor can be synthesized. For our study, we assume that a precursor similar to CHP used by Bieri \textit{et al.}\cite{bieri2009porous} can be synthesized, the only difference being that one of the inner carbons of each molecule is replaced by a nitrogen or boron impurity. We study the electronic properties of these structures using density functional theory (DFT) and employ a tight-binding (TB) parameterization to study the case of more dilute doping. Additionally, a Green's function formalism is used to determine the activation energy for isolated dopants at a low computational cost. To our knowledge, there has been no theoretical work on doped GALs, and we thus report the first theoretical evidence of p- and n-type GAL semiconductors.

\section{Theory and methods}
\label{sec:theory}

\begin{figure}[tb]
	\centering
	\includegraphics[width=0.98\columnwidth]{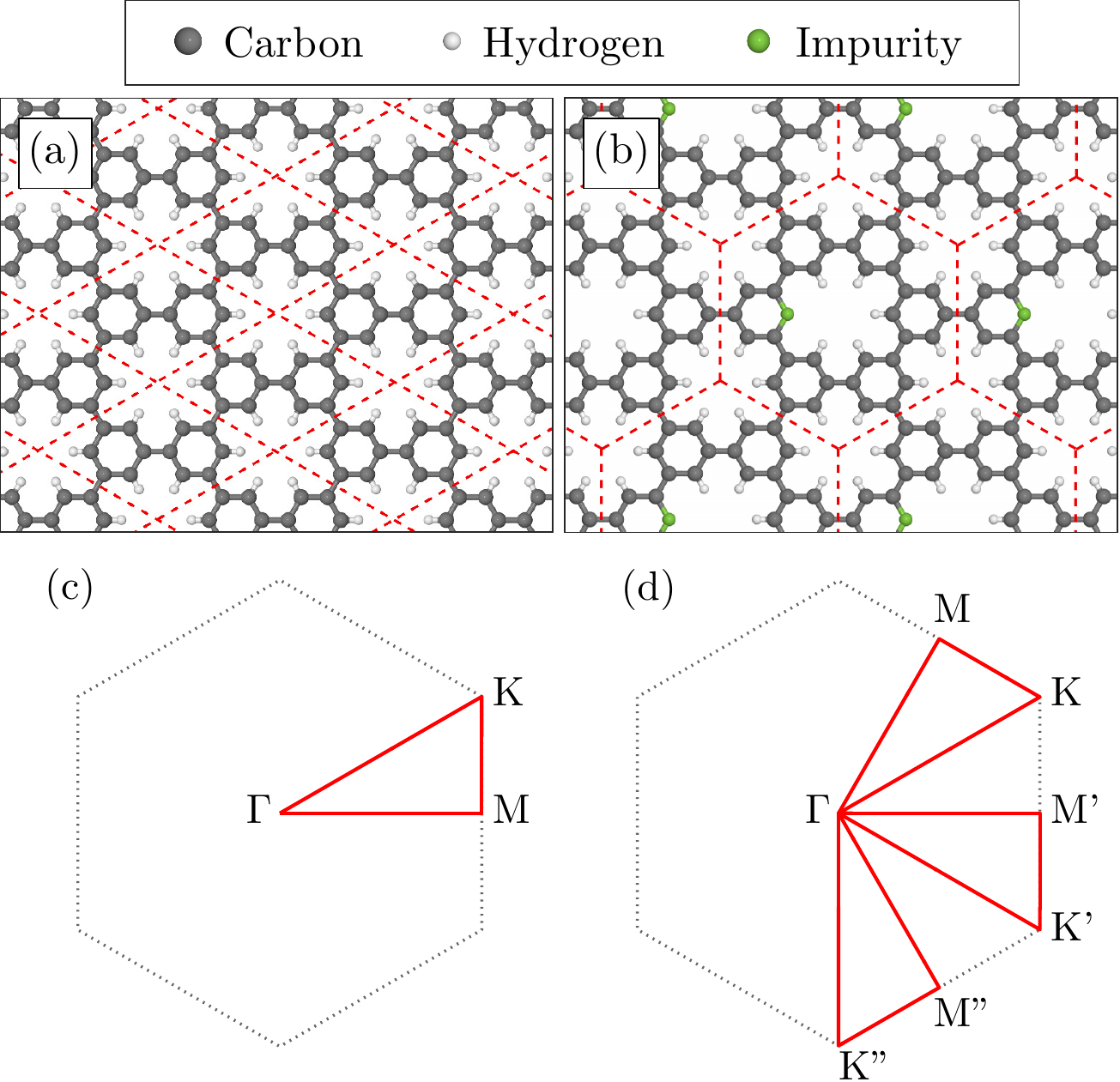}
	\caption{Structural unit cells and corresponding Brillouin zones of the GALs studied in this paper. (a) Unit cell of the pristine system and (c) its Brillouin zone and \mbox{$k$-path} for the band structure. (b) Unit cell for a doped system (in the case of no hydrogen termination at the impurity) with (d) the \mbox{$k$-path} for this structure.}
	\label{fig:fig1}
\end{figure}

The atomic structure of the pristine GAL used in our study is shown in Fig.~\ref{fig:fig1}a, where the dashed red lines mark the primitive unit cells. The properties of GALs have been studied theoretically by several groups.\cite{pedersen2008graphene,furst2009electronic,petersen2009quasiparticle,vanevic2009character,petersen2010clar,dvorak2013bandgap,trolle2013large,yuan2013electronic,brun2014electronic,thomsen2014dirac} In the notation in Ref.~\citenum{petersen2010clar}, the one synthesized by Bieri \textit{et al.} is a rotated GAL (RGAL). It turns out that two thirds of RGALs are semimetals while every third is a semiconductor. Petersen \textit{et al.}\cite{petersen2010clar} have presented a rule based on structural parameters determining if an RGAL is a semimetal or semiconductor and, according to their rule, the antidot lattice described here is semiconducting. Here, we will not study doping in other types of antidot lattices than the experimentally realized one in Fig.~\ref{fig:fig1}a. Therefore, we refer to this type of antidot lattice simply as GAL through the rest of the paper. 

We construct the doped systems from modified CHP molecules, where one carbon atom on the inner edge of the molecule is replaced with either boron or nitrogen. We choose the edge site, as this has been shown to be the most stable site for doping in GNRs.\cite{huang2007making,li2009spin,wang2011selective} An example of the structure for this unit cell is shown in Fig.~\ref{fig:fig1}b. The figure also shows the Brillouin zones and corresponding band structure \mbox{$k$-paths} for both the pristine and doped systems. Because of broken symmetry in the unit cell containing an impurity, the route for the band structure is different than for the pristine system. We place the impurity at an edge site and vary the hydrogen termination between zero, one and two hydrogens at the impurity. Previously, Huang \textit{et al.}\cite{huang2009first} have made theoretical studies of boron and nitrogen doping at graphene edges and shown that the favorable termination for edge doping is one hydrogen (pyridinium-like) both for boron and nitrogen doping. However, Wang \textit{et al.}\cite{wang2011selective} have shown that the most stable configuration may be changed to no hydrogen at the impurity (pyridine-like) by varying the ratio between monohydrogenated and dihydrogenated edge carbon. This suggests that the synthesis may be controlled to yield different degrees of hydrogenation at the impurity, for which reason we choose to study all three kinds of hydrogen termination. Doped structures are studied in the fully ordered configuration, meaning that all precursor molecules are oriented in the same direction. Systems with more dilute doping are also studied, for which some molecules are left undoped. We realize that the orientation would be random for practical synthesis, but we focus on ordered cases to keep the computational cost manageable. All structures are planar except for those with dihydrogenated impurities, where only the two hydrogen atoms on the impurity are out of the plane. 

The pristine and fully ordered doped GALs are studied via DFT using the \textsc{abinit} package,\cite{gonze2002first,gonze2009abinit,bottin2008large,torrent2008implementation} in which a plane wave basis set is used to expand the wave function. We use the Perdew-Burke-Ernzerhof generalized gradient approximation (PBE-GGA) functional\cite{perdew1996generalized} and the projector augmented wave (PAW) method\cite{kresse1999ultrasoft} to solve for the eigenstates of the systems. We use a plane-wave cutoff energy of nearly 600~eV together with an $11 \! \times \! 11 \! \times \! 1$ Monkhorst-Pack \mbox{$k$-grid} sampling. The distance between the layers is 10~Å in order to decouple them electronically, and we use a fairly low Fermi smearing of 68~meV. We perform full structural relaxation of all unit cells before calculating band structures. The structures are relaxed until the maximum force is less than 2.6~meV/Å. We have found that these parameters provide sufficient convergence together with a tolerable computational effort. 

In order to investigate the effects of more dilute doping, we employ a $\pi$-orbital TB model to describe the system, meaning that we concentrate in the electronic processes arising from hopping between the $p_z$ orbitals at each carbon/impurity atom, and disregard bands arising from other orbitals further removed in energy from the Fermi level. The Hamiltonian of the pristine system is given by

\begin{equation}
\mathbf{H_0} = \sum_{i} \varepsilon_p | i \rangle \langle i | + \sum_{i,j} t_{ij} | i \rangle \langle j |, 
\label{eq:Hamiltonian}
\end{equation}

\noindent where $\varepsilon_p$ is the carbon on-site energy and $t_{ij}$ is the hopping integral between atoms $i$ and $j$. We include interactions up to third-nearest neighbors and allow for non-orthogonality in the overlap matrix~$\mathbf{S}$. The impurity is modeled solely by shifting the on-site potential on the impurity with respect to $\varepsilon_p$. The impurity Hamiltonian, which must be added to Eq.~\ref{eq:Hamiltonian}, then becomes $\mathbf{H_1} = \Delta | l \rangle \langle l |$, where the impurity is located at site $l$ and $\Delta$ is the shift of the on-site potential. Other reports include a change in the hopping integral between the impurity and up to its third-nearest neighbors.\cite{lherbier2013electronic,lherbier2015charge} However, we find that this only changes the fit marginally.

\begin{figure}[tb]
	\centering
	\includegraphics[width=0.98\columnwidth]{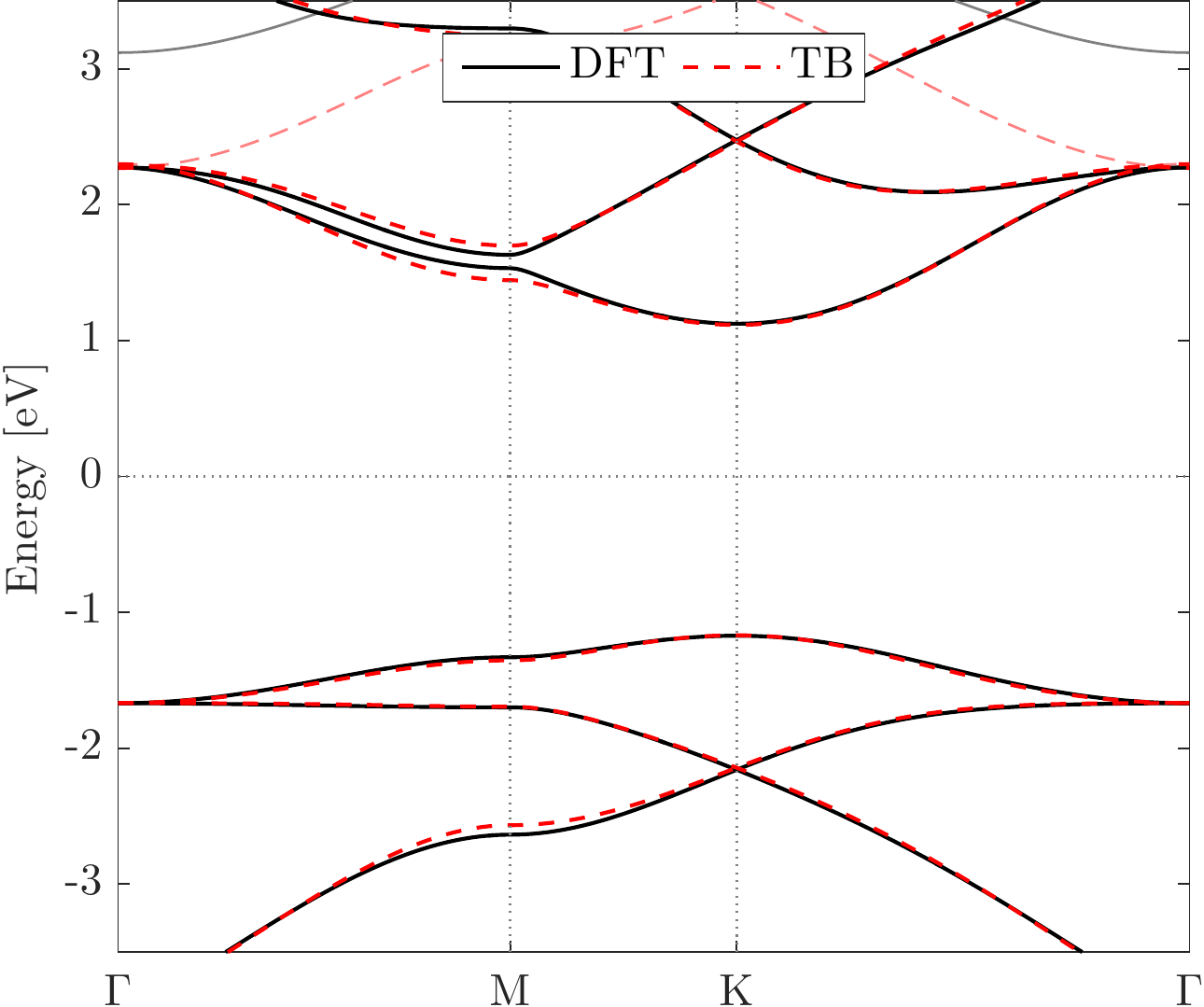}
	\begin{tabular}{c c c c c c c} 
		\vspace{1mm}\\
		\hline
		$\varepsilon_p$ & $t_1$ & $t_2$ & $t_3$ & $s_1$ & $s_2$ & $s_3$
		\\ [0.5ex] 
		\hline 
		-2.18 & -2.58 & -1.11 & -0.29 & 0.36 & 0.08 & 0.07 \\ 
		\hline 
	\end{tabular}
	\caption{Band structure of the pristine GAL shown in Fig.~\ref{fig:fig1}a, calculated using DFT. The best third-nearest neighbor non-orthogonal TB parameterization is also shown. Full colors/lines show the bands used for the TB fit, while weak colors/lines show the rest of the band structures. The TB parameters are listed in the table, where the on-site energy and hopping integrals are in units of eV.}
	\label{fig:fig2}
\end{figure}

\begin{figure*}[tb]
	\centering
	\includegraphics[width=0.96\textwidth]{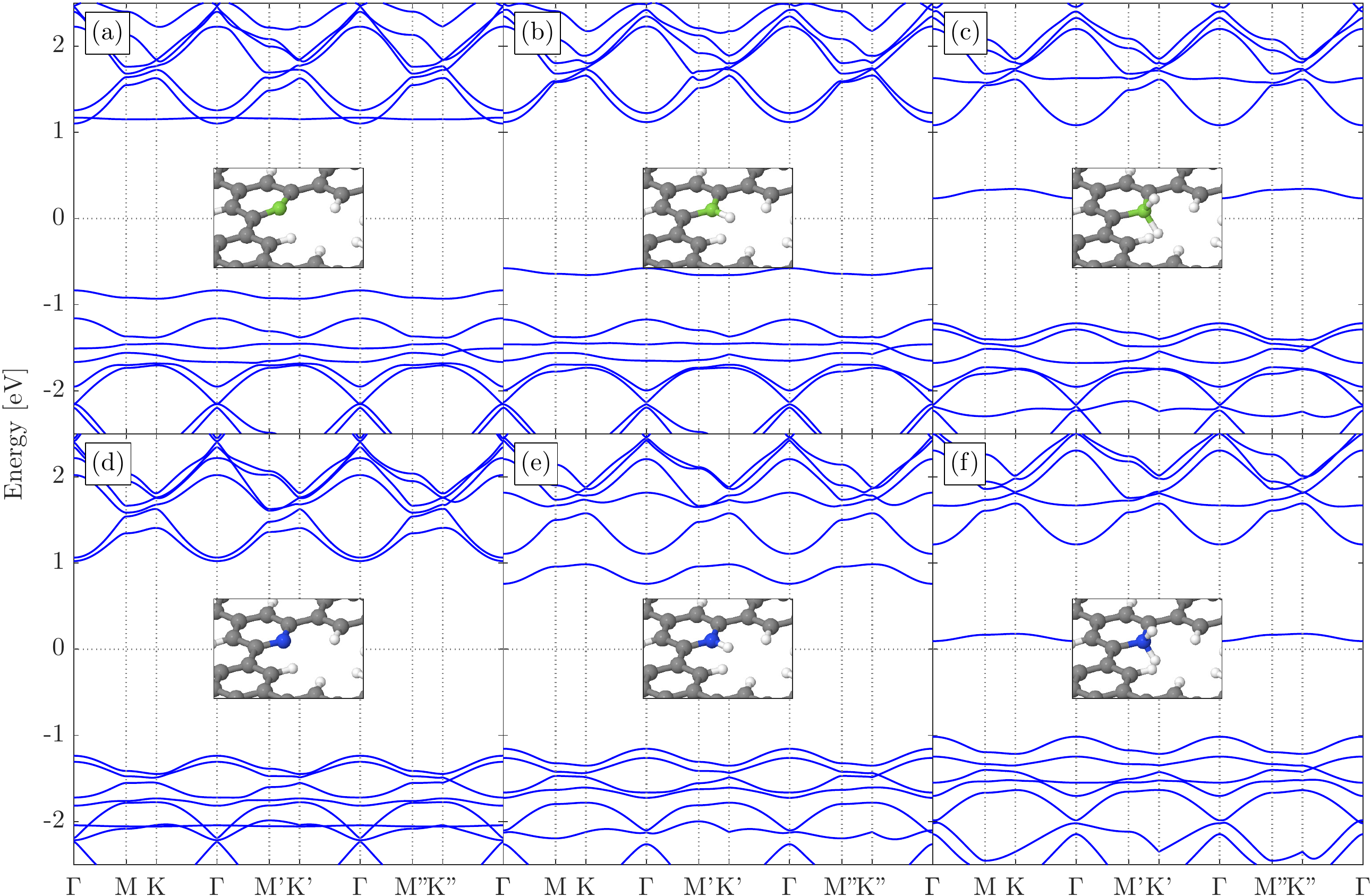}
	\caption{DFT band structures of boron or nitrogen doped GALs for different hydrogen termination on the impurity. Panels (a)-(c) show boron doping and (d)-(f) show nitrogen doping, in both cases terminated by zero, one and two hydrogen atoms at the impurity, respectively.}
	\label{fig:fig3}
\end{figure*}

We begin by calculating the band structure of the pristine system using DFT and obtaining the optimal TB parameterization. The fit is carried out using the three valence and three conduction bands closest to the Fermi level. Figure~\ref{fig:fig2} shows that the DFT band structure can be fitted with excellent agreement by this TB parameterization. We find it necessary to include third-nearest neighbors in a non-orthogonal model for the fit to be in good agreement with DFT. The parameters for the TB model are listed in the table in Fig.~\ref{fig:fig2}. Here, subscripts 1, 2, and 3 denote nearest, second-nearest and third-nearest neighbors, respectively. The structure is a semiconductor, consistent with the rule presented in Ref.~\citenum{petersen2010clar}, and has a rather large band gap of 2.30~eV. This is good agreement with previous DFT calculations for the same structure, i.e. 2.34 eV\cite{du2010multifunctional} and 2.48 eV.\cite{li2010two} For reference, we have also fitted the band structure of pristine graphene to this TB model and, again, find excellent agreement. Moreover, the obtained parameters are in good agreement with those reported by Grüneis \textit{et al.}\cite{gruneis2008tight}

Having an accurate TB parametrization of the electronic structure allows us to efficiently employ a Green's function formalism to analyze the doping level in the case of completely isolated dopants. This formalism for non-orthogonal models was previously developed and used to describe bulk and edge doping in graphene, see Refs.~\citenum{pedersen2013self} and \citenum{pedersen2015self} for further details on the derivation. The theory shows that modeling the impurity by only adjusting its on-site potential yields the following particularly simple expression for the impurity perturbed Green's function at lattice site $l$:

\begin{equation}
\mathbf{G}_{ll}(z) = \frac{\mathbf{G}_{ll}^0(z)}{1-\Delta \widetilde{\mathbf{G}}_{ll}^0(z)},
\label{eq:LDOS}
\end{equation}

\noindent where the Green's functions are given by $\mathbf{G}_0(z) = (z-\mathbf{S}^{-1}\mathbf{H}_0)^{-1}$ and $\widetilde{\mathbf{G}}_0(z) = (z\mathbf{S}-\mathbf{H}_0)^{-1}$. For a semiconductor, the doping level shows up as a pole contribution in the band gap of the impurity local density of states (LDOS). In the limit of vanishing broadening, this approaches a Dirac delta function. The energy of this state, i.e.\ the doping level, may be evaluated in a simple manner by considering Eq.~\ref{eq:LDOS}. The impurity LDOS is given by $L(\omega) = -\pi^{-1} \text{Im}\left\{ \mathbf{G}_{ll}(\omega) \right\}$, which diverges when $\text{Re} \big\{ \widetilde{\mathbf{G}}_{ll}^0(z) \big\} = 1/\Delta$ and $\text{Im} \big\{ \widetilde{\mathbf{G}}_{ll}^0(z) \big\} = 0$ are both satisfied. This means that the doping level may be determined by evaluating the Green's function, assuming the impurity on-site energy shift is known.

\section{Results}

We now proceed to study the effect of replacing one edge carbon in the unit cell with either a boron or nitrogen atom as illustrated in Fig.~\ref{fig:fig1}b. Figure~\ref{fig:fig3} shows DFT band structures for both boron and nitrogen doped GALs with different hydrogen termination. From the top panels, we see that, as expected, boron doping introduces an acceptor level near the highest valence band, which moves closer to the conduction bands as the number of hydrogen atoms on the impurity increases. In the case of two hydrogen atoms, the doping level has even moved across the Fermi level of the pristine structure. Similarly, nitrogen doping introduces a donor level close to the lowest conduction band which moves towards the valence bands as the number of hydrogen atoms at the impurity increases. In the case of no impurity hydrogenation, the doping level is very close to the conduction band edge. Furthermore, we note that the remaining band structure does not change appreciably. 

\begin{figure}[t]
	\centering
	\includegraphics[width=0.98\columnwidth]{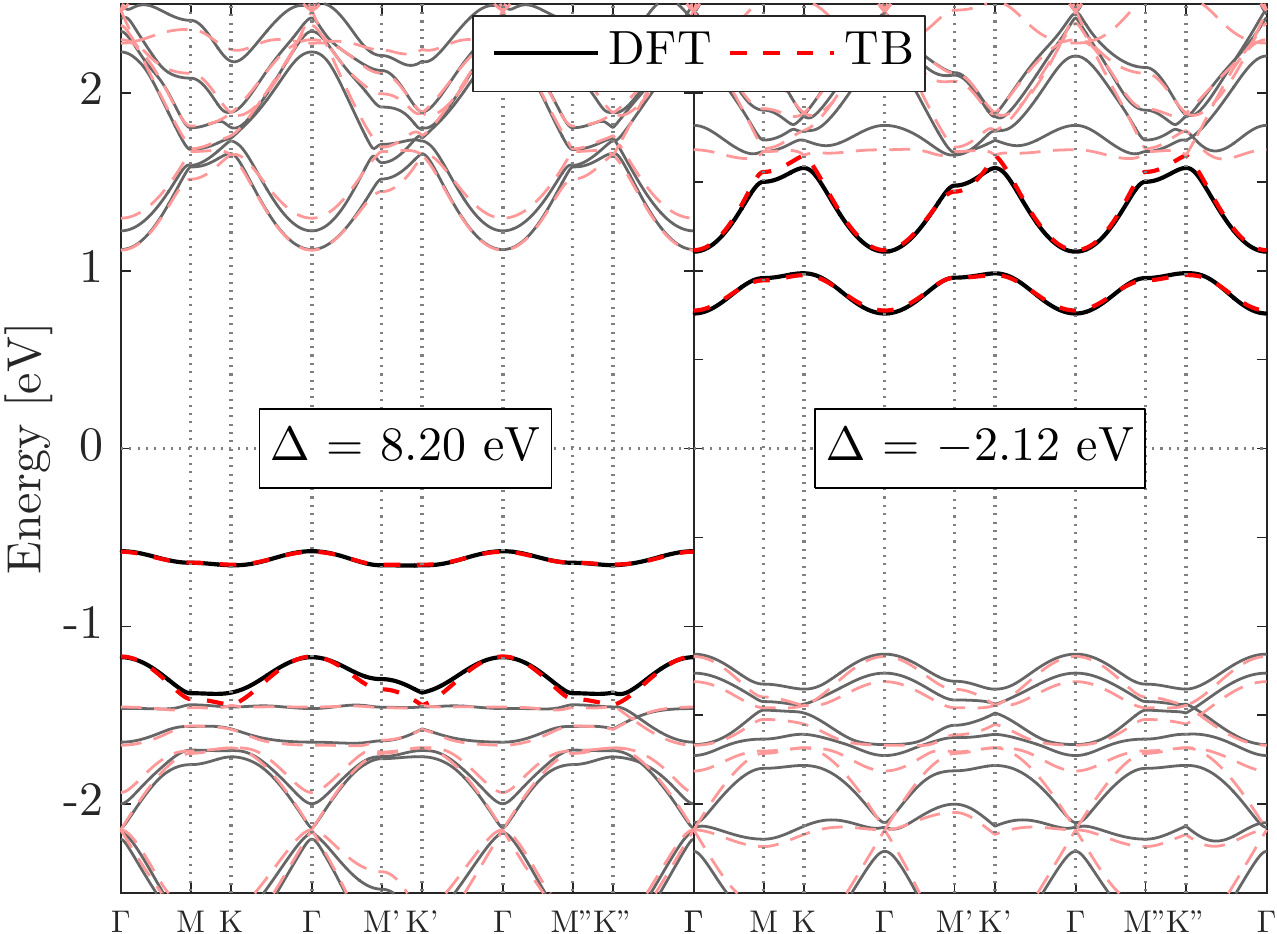}
	\caption{DFT and fitted TB band structures of boron (left) and nitrogen (right) doped GALs, both terminated by one hydrogen atom at the impurity. Full colors/lines show the bands used for the TB fit, while weak colors/lines show the rest of the band structure.}
	\label{fig:fig4}
\end{figure}

We use the parameterization of the pristine system as the basis for the TB model describing the perturbed systems. In our effort to make a good, yet simple model for the perturbed systems, we model the impurity by only adjusting the on-site element at the impurity site, making $\Delta$ the only fitting parameter for the perturbed systems. Examples of fits for boron and nitrogen doping, both terminated by one hydrogen (corresponding to Figs.~\ref{fig:fig3}b and \ref{fig:fig3}e, respectively), are shown in Fig.~\ref{fig:fig4}. In the fit, we include only the two valence (conduction) bands closest to the Fermi level for boron (nitrogen) doping. The fits demonstrate that such a simple model still yields a very good description of the bands in the vicinity of the Fermi energy for both types of doping. The fitted values of $\Delta$ are listed in Table~\ref{tab:delta_fit}. Only for boron terminated by two hydrogen atoms were we unable to obtain a satisfactory fit. 

\begin{table}[b] 
	\begin{tabular}{l | c c c}
		Impurity hydrogenation   & 0      & 1      & 2 \\
		\hline
		Boron $\Delta$ [eV]      & 3.22  & 8.20  & - \\
		Nitrogen $\Delta$ [eV]   & -0.88 & -2.12 & -6.94 \\
	\end{tabular}
	\caption{Fitted values of $\Delta$ for boron and nitrogen doping and for different hydrogen terminations at the impurity.}
	\label{tab:delta_fit}
\end{table} 

\begin{figure}[tb]
	\centering
	\includegraphics[width=0.98\columnwidth]{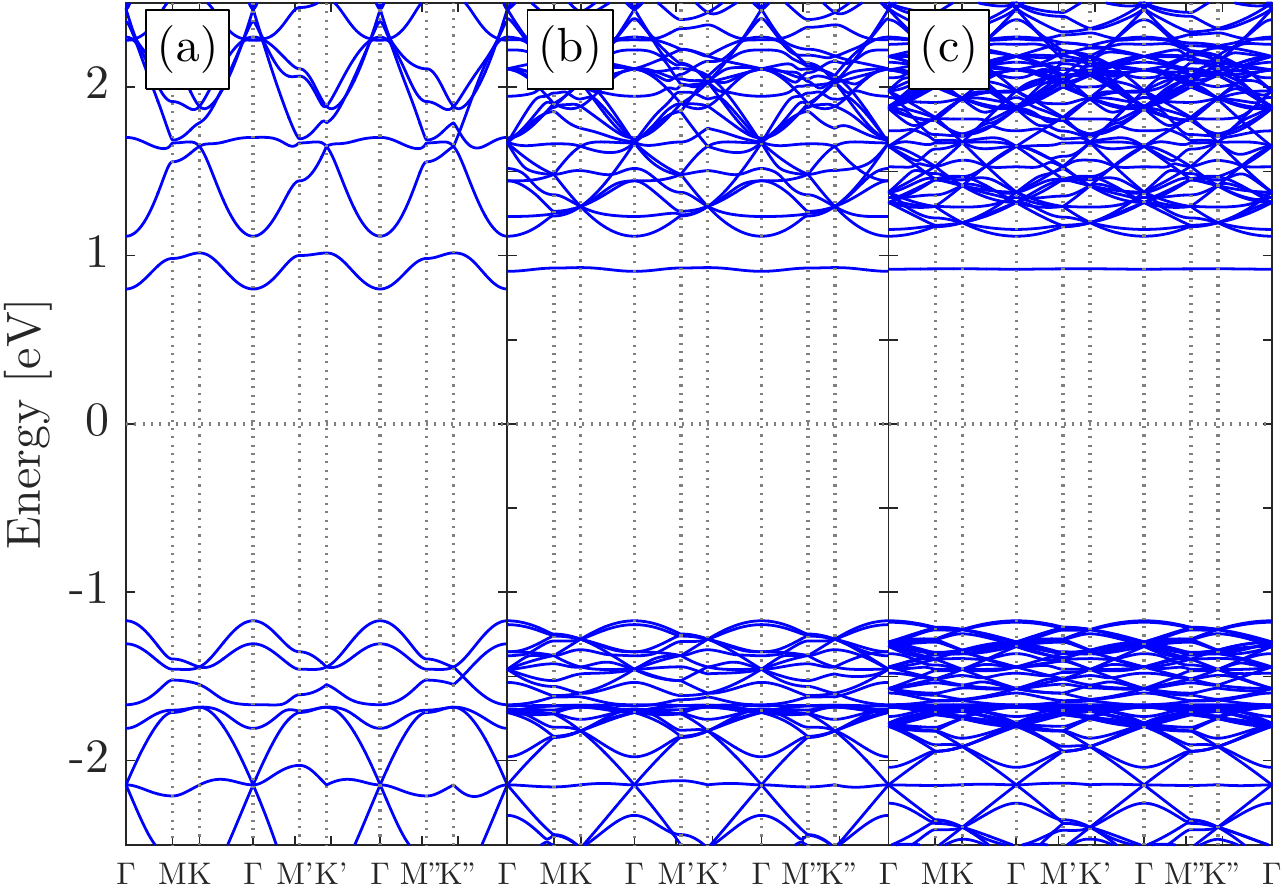}
	\caption{TB band structures of supercells containing (a)~1$\times$1, (b)~2$\times$2 and (c)~3$\times$3 precursor unit cells (see Fig.~\ref{fig:fig1}b) with only one impurity per supercell. The impurity is modeled using $\Delta = -2$~eV.}
	\label{fig:fig5}
\end{figure}

The unit cells in Fig.~\ref{fig:fig1}b used for the DFT calculations are relatively small and place the impurities only 12.8~Å apart. This is also evident from the significant dispersion of the impurity bands seen in the band structures of Fig.~\ref{fig:fig3}. The TB parametrization allows us to go comfortably beyond that, and calculate band structures for supercells consisting of $N \! \times \! N$ precursor unit cells, where each supercell contains only one impurity. Specifically, we study the doping level as the doping concentration decreases. Figure~\ref{fig:fig5} shows band structures of $1 \times 1$, $2 \times 2$ and $3 \times 3$ supercells containing only one impurity for which $\Delta = -2$~eV. It is clear that the mid-gap impurity band becomes increasingly flatter as the cell size increases. In this case, a $3 \times 3$ supercell is enough to get a nearly dispersionless impurity band. However, for values of $\Delta$ closer to zero, the convergence is worse and a much larger cell is required. This is not surprising because the extent of the wavefunction associated with these impurity levels is determined by their distance to the nearest band and, consequently, shallower donors/acceptors tend to be hybridized over larger spatial scales. Note that the most striking impact of changing the supercell size takes place in the impurity band. The denser nature of the conduction and valence bands as we go from panels (a) to (c) in Fig. 5 is simply due to band folding, as the unit cell size is increased.

\begin{figure}[b]
	\centering
	\includegraphics[width=0.98\columnwidth]{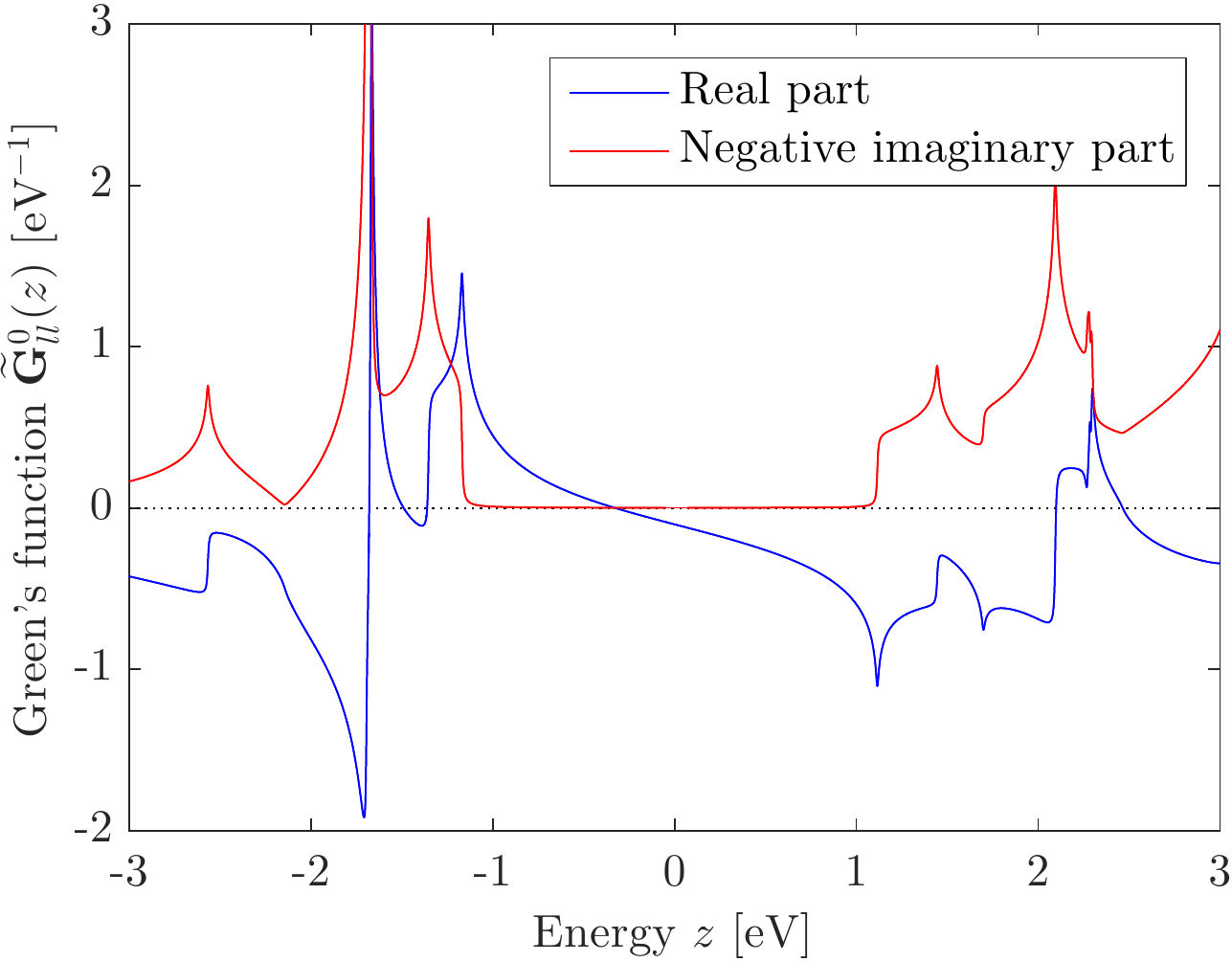}
	\caption{Green's function for the impurity site of the GAL with a broadening of 5~meV.}
	\label{fig:fig6}
\end{figure}

\begin{figure}[t]
	\centering
	\includegraphics[width=0.98\columnwidth]{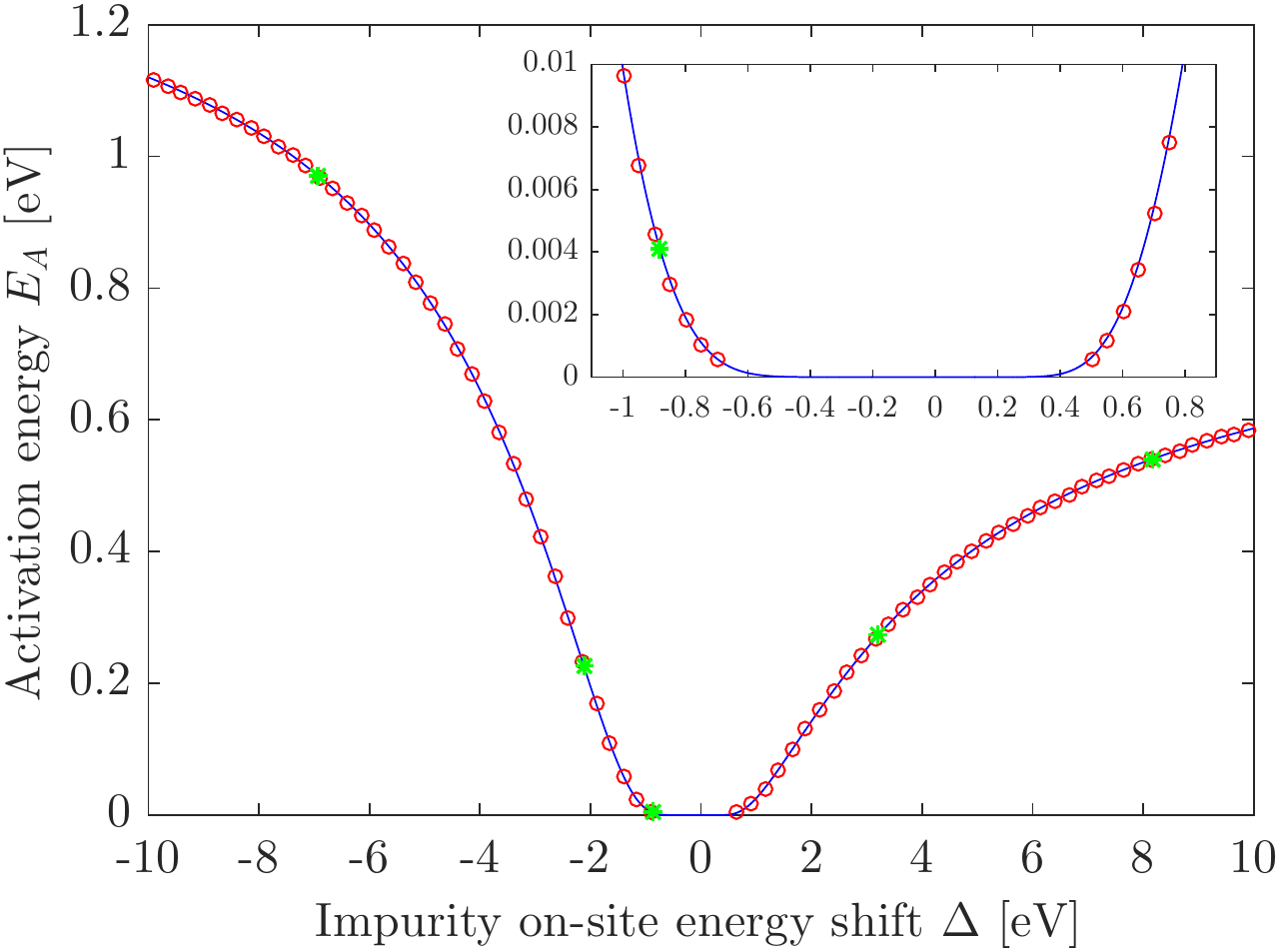}
	\caption{Activation energy for different values of the impurity on-site shift $\Delta$, where the Green's function method (solid blue line) is compared with the supercell band structure method (red circles). The actual values of $\Delta$ obtained ab initio and listed in Table~\ref{tab:delta_fit} are also shown (green asterisks).}
	\label{fig:fig7}
\end{figure}

Once the doping level in the TB band structure is sufficiently flat, we are able to determine the activation energy of the system. However, we may also calculate the doping level of completely isolated impurities using the Green's function technique described in Sec.~\ref{sec:theory}. The local Green's function at the impurity site is shown in Fig.~\ref{fig:fig6}, where the band gap is clearly seen as the region where the imaginary part is zero. The real part of the Green's function is used to calculate the activation energy. For negative values of $\Delta$ (corresponding to n-type doping), the activation energy is given by $E_A = E_c-E_d$, while it is $E_A = E_d-E_v$ for positive values (corresponding to p-type doping). Here, $E_d$ refers to the energy of the doping level while $E_v$ and $E_c$ are the highest (lowest) energy of the valence (conduction) band, respectively. The doping level is found by solving $\text{Re} \big\{ \widetilde{\mathbf{G}}_{ll}^0(z) \big\} = 1/\Delta$ for the energy~$z$ within the band gap region. This calculation is very fast when a converged Green's function is provided. Due to the relatively simple structure of the pristine GAL, calculating the Green's function is computationally straightforward. The activation energy as a function of $\Delta$ is shown in Fig.~\ref{fig:fig7}, where the fitted values from Table~\ref{tab:delta_fit} are marked with green asterisks. The activation energies from the supercell band structures are also shown, although we emphasize that this latter method is much more computationally demanding, as several band structures have to be calculated for each value of $\Delta$. Furthermore, a dispersionless impurity level requires very large supercells for values of $\Delta$ close to zero, making them extremely time consuming. The results from the supercell band structures (red circles) are in excellent agreement with the curve obtained using the Green's function method, thus verifying the result. Because of the slow convergence for $\Delta$ close to zero, the supercell method has only been used for values outside the $\pm 0.5$~eV regime. This is shown in the inset of the figure, where the agreement is seen to continue for all the values provided. We also note that nitrogen doping with no impurity hydrogenation results in a very low activation energy of 4.1~meV. The above results point to the advantage of using the Green's function method even when the system is not in the strictly dilute limit. Its application is not limited in any way to the specific structure we considered here, and is applicable to any system of dilute impurities in a crystal lattice, such as other antidot lattice geometries, provided an accurate TB model is available. We are convinced that our analysis of the properties of doped GALs will be useful for future studies of electronic and transport properties of junctions in graphene nanostructures.


\section{Conclusion}

We have studied the effect of substituting an edge carbon atom in a GAL with either boron or nitrogen. By means of DFT, we calculate electronic band structures for GALs where the impurity is terminated by zero, one or two hydrogen atoms. We perform TB parameterizations describing both the pristine and doped systems with high accuracy, which are used together with a Green's function method to study more dilute doping. Boron doping introduces an acceptor level near the valence band edge, which moves towards the conduction bands as the hydrogenation on the impurity increases. Similarly, nitrogen introduces a donor level near the conduction band, which moves towards the valence bands with increasing hydrogenation. This indicates that the properties of doped GALs may be tuned, provided the impurity hydrogenation is controllable during synthesis, as suggested in Ref.~\citenum{wang2011selective}. Our work is the first step on the way to understanding doping in GALs. We believe that the parameterizations reported here, together with the activation energy analysis are useful tools for further studies of this and other types of doped GALs.

\section*{Acknowledgments}

SJB and TGP gratefully acknowledge the financial support from the Center for Nanostructured Graphene (Project No. DNRF103) financed by the Danish National Research Foundation and from the QUSCOPE project financed by the Villum Foundation. SJB further acknowledges the hospitality and support of the NUS Centre for Advanced 2D Materials, where part of this research was conducted. VMP was partly supported by the National Research Foundation (Singapore) under its Medium Size Centre programme.


\bibliography{literature}

\end{document}